\documentclass[12pt]{article}
\usepackage{amssymb}
\usepackage{pifont}
\usepackage{mathrsfs}
\usepackage{amsfonts}
\usepackage{color}
\usepackage{footmisc}
\usepackage[numbers,sort&compress]{natbib}
\usepackage[bookmarksopen=true, bookmarksopenlevel=\maxdimen,
bookmarksnumbered=true,colorlinks,linkcolor=blue,
citecolor=blue,urlcolor=blue]{hyperref}

\newcommand{\omits}[1]{}

\begin{document}

\begin{center}
{\bf \LARGE Energy, momentum and angular momentum\bigskip

conservations in de Sitter gravity}
\bigskip\bigskip

{\large Jia-An Lu$^{a}$\footnote{Email: ljagdgz@163.com}}
\bigskip

$^a$ School of Physics and Astronomy, Sun Yat-sen
University,\\ Guangzhou 510275, China
\bigskip

\begin{abstract}
In de Sitter (dS) gravity, where gravity is a gauge field introduced to realize the
local dS invariance of the matter field, two kinds of conservation laws are derived.
The first kind is a differential equation for a dS-covariant current, which unites the canonical
energy-momentum (EM) and angular momentum (AM) tensors. The second kind presents a dS-invariant
current which is conserved in the sense that its torsion-free divergence vanishes.
The dS-invariant current unites the total (matter plus gravity) EM and AM currents.
It is well known that the AM current contains an inherent part, called the spin current.
Here it is shown that the EM tensor also contains an inherent part, which might be observed
by its contribution to the deviation of the dust particle's world line from a geodesic.
All the results are compared to the ordinary Lorentz gravity.
\end{abstract}
\end{center}

\quad {\small PACS numbers: 04.50.Kd, 04.20.Cv, 04.20.Fy}

\quad {\small Key words: de Sitter gravity, conservation law, inherent energy-momentum tensor}

\section{Introduction}

It follows from the Noether's theorem that each one-parameter group of symmetric transformations
gives rise to a conservation law \cite{Noether,Peskin}. Especially, each one-parameter group of
translations or Lorentz transformations in special relativity (SR) results in a Lorentz-invariant
conserved current. By eliminating the generators of the one-parameter groups,
the conservation laws lead to a set of differential equations for the canonical energy-momentum (EM)
and spin tensors. The Lorentz-covariant energy, momentum and angular momentum (EMA) currents can be
constructed out of these two tensors.

It is rather nontrivial to generalize these results to the gravitational theories.
A generalization can be performed in Lorentz gravity, where gravity is represented
by the metric-compatible connection and tetrad 1-forms which are introduced to realize the
local Lorentz invariance of the matter field \cite{Utiyama}.
In Lorentz gravity, it is shown that the diffeomorphism and Lorentz symmetries lead
to a set of differential equations for the EM and spin tensors \cite{Kibble,Hehl76,Hehl95}.
Note that the total (matter plus gravity) EM and spin tensors
vanish due to the gravitational field equations.
On the other hand, provided the gravitational field equations are satisfied,
each one-parameter group of translations or Lorentz rotations results in a Lorentz-invariant
current, which is conserved in the sense that its torsion-free divergence vanishes
\cite{Obukhov06,Obukhov08}. The Lorentz-invariant currents can be used to define the total
EMA currents in Lorentz gravity.

Note that in Lorentz gravity, gravity is not a gauge field, in
the sense that it is not an Ehresmann connection of some principal fiber bundle.
On the other hand, there exists the de Sitter (dS) gravity \cite{Stelle,PoI,Guo07,Lu13,Lu14},
where gravity is described by an Ehresmann connection which is introduced to realize the
local dS invariance of the matter field.
The dS gravity is well motivated for some cosmological reasons.
Firstly, the observed cosmological constant may be related to that of the internal dS space,
which is a characteristic structure in dS gravity.
Moreover, an interesting investigation shows that \cite{Lu14,Lu14-b},
the dS symmetry together with a Kaluza--Klein-type ansatz can pick out the only one model that is
free of the big-bang singularity in the Robertson--Walker universe filled with a spin fluid \cite{Kuchowicz},
among the $R+\beta S^{abc}S_{abc}$ models of gravity \cite{Vignolo}, where $R$ is the scalar curvature,
$\beta$ is a parameter, and $S_{abc}$ denotes the torsion tensor.

In this paper, the EMA conservation laws are generalized to dS gravity.
The result consists of two kinds of conservation laws.
Firstly, it is shown that the diffeomoriphism and dS symmetries lead to
a differential equation for a dS-covariant current, which
unites the EM and spin tensors. Secondly, provided the gravitational
field equation is satisfied, each one-parameter group of dS rotations result in
a dS-invariant current, which is conserved in the sense that its torsion-free divergence vanishes.
The dS-invariant current can be used to define the total (matter plus gravity) EMA currents
in dS gravity. In the analysis of the first kind conservation law,
it is found that the EM tensor contains an inherent part, just like the fact that
the angular momentum (AM) current contains an inherent part (the spin current).
Also, the dust particle's world line is studied, which deviates from a geodesic for
two reasons. The first is the existence of the spin tensor, while the second is the
existence of the inherent EM tensor discovered here.

The paper is arranged as follows. In section 2, the dS gravity is briefly introduced.
In sections 3--4, two kinds of EMA conservation laws are derived in dS gravity, and compared to those in
Lorentz gravity. Moreover, in section 3, an inherent EM tensor is predicted and its impact on the
dust particle's world line is discussed. Finally, we end with some remarks in the last section.

\section{de Sitter gravity}

The dS gravity is a gauge theory of gravity with local dS invariance.
It can \cite{PoI,Guo07} be seen as
the localization of dS SR \cite{Dirac,Gursey,Fantappie,Pessa,Guo-MPL,Guo-PLA,Yan,Xu,Aldrovandi,Lu15},
which is a hypothetical theory with global dS invariance. Consider a matter field $\psi$
on a dS space ${\cal M}_l$, with the action integral $S_M$ and Lagrangian function
$\mathscr{L}_M$ as follows:
\begin{equation}
S_M=\int_\Omega\mathscr{L}_M\epsilon,\ \
\mathscr{L}_M=\mathscr{L}_M(\psi, d_a\psi,c.c., \xi^A, d_a\xi^A),
\end{equation}
where $\Omega$ is an arbitrary domain of ${\cal M}_l$, $\epsilon$ is a metric-compatible volume element,
$d_a$ is the ordinary exterior derivative, $a$ is an abstract index \cite{Wald,Liang} which can be
changed into any tetrad or coordinate index by taking the corresponding component,
$c.c.$ denotes the complex conjugate of the former quantities, $\xi^A$ are 5-dimensional (5d) Minkowski
coordinates which satisfy $\eta_{AB}\xi^A\xi^B=l^2$ on ${\cal M}_l$, $A=0,1,\cdots4$ are dS indices,
$\eta_{AB}={\rm diag}(-1,1, \cdots 1)$, and $l$ is a constant with dimension of length.
Suppose that the action integral $S_M$ is globally dS invariant, i.e., it is invariant under the
transformation:
\begin{equation}\label{dST}
\psi\rightarrow T(g)\psi,\ \ \xi^A\rightarrow g^A{}_B\xi^B,
\end{equation}
where $g=g^A{}_B$ is an element of the dS group $O(1,4)$, and $T$ is the representation of
$O(1,4)$ associated with the matter field $\psi$. Note that the dS metric
\begin{equation}
g_{ab}=\eta_{AB}(d_a\xi^A)(d_a\xi^B)
\end{equation}
is globally dS invariant, so as $\epsilon$. The localization of the above theory is to
introduce a gauge field (the gravitational field) into $S_M$, such that
$S_M$ is invariant under the localized version of the transformation (\ref{dST})
(obtained by changing $g=g^A{}_B$ into an $O(1,4)$-valued function),
supplemented by the transformation for the gauge field.
The result is as follows. The dS space ${\cal M}_l$
is changed to be a Riemann--Cartan (RC) spacetime $\cal M$, and the Lagrangian function
of the matter field is changed to be
\begin{equation}
\mathscr{L}_M=\mathscr{L}_M(\psi, D_a\psi,c.c., \xi^A, D_a\xi^A),
\end{equation}
where
\begin{equation}\label{dpsi-dS}
D_a\psi=d_a\psi+T_A{}^B\Omega^A{}_{Ba}\psi,
\end{equation}
\begin{equation}\label{Dxi}
D_a\xi^A=d_a\xi^A+\Omega^A{}_{Ba}\xi^B,
\end{equation}
$T_A{}^B$ are representations of the dS generators, $\Omega^A{}_{Ba}$
is the gravitational field as an $o(1,4)$-valued 1-form, $o(1,4)$ is the dS
algebra, and $\xi^A$ become local functions which satisfy
$\eta_{AB}\xi^A\xi^B=l^2$. The action integral $S_M$ is locally dS invariant,
i.e., invariant under the combination of the localized version of transformation
(\ref{dST}) and
\begin{equation}
\Omega^A{}_{Ba}\rightarrow
g^A{}_C\Omega^C{}_{Da}(g^{-1})^D{}_B
+g^A{}_C d_a(g^{-1})^C{}_B,
\end{equation}
where $g^A{}_C$ is an $O(1,4)$-valued function. Note that the metric \cite{Stelle}
\begin{equation}\label{metric}
g_{ab}=\eta_{AB}(D_a\xi^A)(D_a\xi^B)
\end{equation}
is locally dS invariant, so as $\epsilon$. The gravitational action integral
$S_G$ and Lagrangian function $\mathscr{L}_G$ are given by
\begin{equation}\label{SG}
S_G=\int_\Omega\mathscr{L}_G~\epsilon,\ \
\mathscr{L}_G=\mathscr{L}_G(\xi^A, D_a\xi^A, {\cal F}^A{}_{Bab}),
\end{equation}
where
\begin{equation}\label{F}
{\cal F}^A{}_{Bab}=d_a\Omega^A{}_{Bb}+\Omega^A{}_{Ca}\wedge\Omega^C{}_{Bb}
\end{equation}
is the curvature of $\Omega^A{}_{Ba}$. The final theory of dS gravity is
described by the total action integral $S=S_M+\kappa S_G$, where $\kappa$
is a coupling constant. It is worth noting that the dS gravity defined here
is different from that described by Cartan geometry \cite{Wise} or nonlinear
realisation \cite{Tseytlin}, where the gravitational field as a Cartan connection
or nonlinear gauge field only performs Lorentz transformations, but not
the full dS transformations.

Here we briefly introduce a suitable mathematical framework for dS gravity \cite{Lu13,Lu14}.
Over any RC spacetime $\cal M$, a principal bundle $\cal P$ can be set up, with the
structure group being $O(1,4)$. An Ehresmann connection $\widetilde{\Omega}$ can be
defined on $\cal P$, such that $\Omega^A{}_{Ba}=(\sigma^*\widetilde{\Omega})^A{}_{Ba}$,
where $\sigma$ is a local section of $\cal P$, and $\sigma^*$ denotes the pullback by $\sigma$.
Let ${\cal Q}$ be a fiber bundle associated to ${\cal P}$, with ${\cal M}_l$ as the typical
fiber. A global section $\Phi$ of ${\cal Q}$ can be defined, such that $\xi^A$ are the
(vertical) 5d Minkowski coordinates of $\Phi$. With these definitions, the curvature tensor
$R^c{}_{dab}$ and torsion tensor $S^c{}_{ab}$ of ${\cal M}$ have the following dS-invariant
expressions \cite{Lu13,Lu14}:
\begin{equation}\label{curvature}
R_{cdab}-(2/l^{2})g_{a[c}g_{d]b}=\mathcal
{F}_{ABab}(D_{c}\xi^{A})(D_{d}\xi^{B}),
\end{equation}
\begin{equation}\label{torsion}
S_{cab}=\mathcal{F}_{ABab}(D_{c}\xi^{A})\xi^{B},
\end{equation}
where $R_{cdab}=g_{ce}R^e{}_{dab}$, $S_{cab}=g_{cd}S^d{}_{ab}$,
and ${\cal F}_{ABab}=\eta_{AC}{\cal F}^C{}_{Bab}$. Let $F$ be the vector space at which $\psi$
is valued, and ${\cal Q}_F$ be a fiber bundle associated to ${\cal P}$, with $F$ as the
typical fiber. A global section $\overline{\psi}$ of ${\cal Q}_F$ can be defined, such that
for any local section $\sigma$ of ${\cal P}$,
$\overline{\psi}=\sigma\cdot\psi=\{(\sigma g,g^{-1}\psi)|g\in O(1,4)\}$.

With the help of Eqs. (\ref{metric}), (\ref{curvature}) and (\ref{torsion}),
it can be shown that the gravitational
Lagrangian function given by Eq. (\ref{SG}) is equivalent to
\begin{equation}\label{LG}
\mathscr{L}_G=\mathscr{L}_G(g_{ab},R^c{}_{dab},S^c{}_{ab}).
\end{equation}
As a simple example, the Einstein--Cartan gravity  \cite{Cartan,Hehl76}
with $\mathscr{L}_G=R$ can be a model
for dS gravity, where $R$ is the scalar curvature. Note that the Lagrangian function
(\ref{LG}) is also Lorentz invariant. To determine which kind of gravity (Lorentz or dS)
it describes one should consider which kind of matter field
it is coupled to. If it is coupled to a
Lorentz-invariant matter field, it describes Lorentz gravity. If it is coupled to a
dS-invariant matter field, it describes dS gravity.

\section{Conservation law A}

Consider a matter field $\psi$ on an RC spacetime $\cal M$, with the dS-invariant action integral
$S_M$ and Lagrangian function $\mathscr{L}_M$ as follows:
\begin{equation}\label{SM}
S_M=\int_\Omega\mathscr{L}_M\epsilon,\ \
\mathscr{L}_M=\mathscr{L}_M(\psi, D_a\psi,c.c., \xi^A, D_a\xi^A).
\end{equation}
The gauge transformations are defined as the bundle isomorphisms on $\cal P$.
Let $\{\widetilde{\phi}_t\}$ be a one-parameter group of gauge transformations, with $t$ as the group parameter.
Then the following transformations can be induced by $\widetilde{\phi}_t$:
\begin{eqnarray}\label{gaugedSM}
\Omega\rightarrow\Omega_t=\phi_t[\Omega],\nonumber \\
\psi\rightarrow\psi_t=T(g_t)\phi_{t*}\psi,\ \
D_a\psi\rightarrow (D_a\psi)_t=T(g_t)\phi_{t*}(D_a\psi),
\nonumber \\
\xi^A\rightarrow \xi_t^A=g_t^A{}_B\phi_{t*}\xi^B,\ \
D_a\xi^A\rightarrow (D_a\xi^A)_t=g_t^A{}_B\phi_{t*}(D_a\xi^B),
\end{eqnarray}
where $\phi_t$ are diffeomorphisms on $\cal M$ induced by $\widetilde{\phi}_t$,
$\phi_{t\ast}$ denote the pushforwards by $\phi_t$,
and $g_t=g_t^A{}_B$ are functions valued at the special dS group $SO(1,4)$,
defined by $\widetilde{\phi}_{t*}\sigma=\sigma g_t$, where $\widetilde{\phi}_{t*}\sigma$
are defined by $(\widetilde{\phi}_{t*}\sigma)(\phi_tx)=\widetilde{\phi}_t(\sigma(x))$,
$\forall x\in{\cal M}$. Let $v^a$ be the generator of $\{\phi_t\}$, $A^A{}_B=\delta g^A{}_B$,
where $\delta=(d/dt)|_{t=0}$. If $\widetilde{\phi}_t$ are dS rotations, then $v^a=0$. If $\widetilde{\phi}_t$
are horizontal, i.e., $\widetilde{\Omega}(d(\widetilde{\phi}_tp)/dt)|_{t=0}=0$, $\forall p\in{\cal P}$, then
\begin{equation}\label{horizontal}
A^A{}_B=-\Omega^A{}_{Ba}v^a.
\end{equation}
Generally, we may assume
\begin{equation}\label{B}
A^A{}_B=B^A{}_B-\Omega^A{}_{Ba}v^a,
\end{equation}
where $B^A{}_B$ is covariant in the following sense:
if $\sigma\rightarrow\sigma g^{-1}$, then
$B^A{}_B\rightarrow g^A{}_C B^C{}_D (g^{-1})^D{}_B$,
where $g=g^A{}_B$ is an arbitrary function valued at $O(1,4)$.
It follows from Eq. (\ref{gaugedSM}) that
\begin{eqnarray}
\delta\psi=B^A{}_BT_A{}^B\psi-v^aD_a\psi,\ \
\delta\xi^A=B^A{}_B\xi^B-v^aD_a\xi^A,\nonumber\\
\delta D_a\psi=B^A{}_BT_A{}^BD_a\psi-v^bD_bD_a\psi-(D_b\psi)\mathring{\nabla}_av^b,\nonumber\\
\delta D_a\xi^A=B^A{}_BD_a\xi^B-v^bD_bD_a\xi^A-(D_b\xi^A)\mathring{\nabla}_av^b,
\end{eqnarray}
where $\mathring{\nabla}_a$ is the metric-compatible and torsion-free derivative.
On account of the invariance of $S_M$ under Eq. (\ref{gaugedSM}),
$\delta\mathscr{L}_M=-v^a\mathring{\nabla}_a\mathscr{L}_M$.
Suppose that the matter field equation is satisfied, then substitution of the above results into
the chain rule for $\delta\mathscr{L}_M$ leads to
\begin{equation}\label{dsigmaG}
\mathring{\nabla}_a\Sigma_b{}^a=-\tau_A{}^{Ba}{\cal F}^A{}_{Bab}
+\frac{\partial\mathscr{L}_M}{\partial\xi^A}D_b\xi^A+
\frac{\partial\mathscr{L}_M}{\partial D_a\xi^A}D_bD_a\xi^A,
\end{equation}
\begin{equation}\label{dtauG}
D_a\tau_{AB}{}^a=-\left(\frac{\partial\mathscr{L}_M}{\partial\xi^{[A}}\xi_{B]}
+\frac{\partial\mathscr{L}_M}{\partial D_a\xi^{[A}}D_a\xi_{B]}\right),
\end{equation}
\begin{equation}\label{dpsi-dxi}
\frac{\partial\mathscr{L}_M}{\partial D_a\psi}D_b\psi+c.c.
+\frac{\partial\mathscr{L}_M}{\partial D_a\xi^A}D_b\xi^A=0,
\end{equation}
where the arbitrariness of $B^A{}_B$, $v^a$ and $\mathring{\nabla}_av^b$ at any given point is used,
\begin{eqnarray}\label{sigmadSG}
\Sigma_b{}^a=-\frac{\partial\mathscr{L}_M}{\partial D_a\psi}D_b\psi+c.c.
+\mathscr{L}_M\delta^a{}_b
\end{eqnarray}
is the orbital EM tensor,
\begin{equation}\label{taudSG}
\tau_A{}^{Ba}=\frac{\partial\mathscr{L}_M}{\partial D_a\psi}T_A{}^B\psi+c.c.
\end{equation}
is the dS-covariant spin current, and
\begin{equation}
D_bD_a\xi^A=\mathring{\nabla}_bD_a\xi^A+\Omega^A{}_{Bb}D_a\xi^B.
\end{equation}
Define the 5d dS-covariant orbital AM current
\begin{equation}\label{sigmaABG}
\Sigma_A{}^{Ba}=\Sigma_b{}^a\eta_{AC}(D^b\xi^{[C})\xi^{B]},
\end{equation}
and the 5d dS-covariant AM current
\begin{equation}\label{VABaG}
V_A{}^{Ba}\equiv\delta S_M/\delta\Omega^A{}_{Ba}=\Sigma_A{}^{Ba}+\tau_A{}^{Ba}.
\end{equation}
In the dS SR limit, $V_A{}^{Ba}$ unites the EMA currents in an inertial coordinate system \cite{Lu15}.
Making use of Eqs. (\ref{dsigmaG})--(\ref{dpsi-dxi}), and the identities
\begin{equation}
(D_a\xi^A)(D^a\xi^B)=\eta^{AB}-\xi^A\xi^B/l^2,
\end{equation}
\begin{equation}
D_aD_b\xi^A=-K^c{}_{ba}D_c\xi^A-g_{ba}\xi^A/l^2,
\end{equation}
where
\begin{equation}
K^c{}_{ab}=\frac12(S^c{}_{ab}+S_{ab}{}^c+S_{ba}{}^c)
\end{equation}
is the contorsion tensor, it can be shown that
\begin{equation}\label{dVG}
D_aV_A{}^{Ba}=V_C{}^{Da}{\cal F}^C{}_{Dba}(D^b\xi_{[A})\xi_{E]}\eta^{EB}.
\end{equation}
We call this result the conservation law A. In the dS SR limit, it becomes
$\mathring{\nabla}_aV_A{}^{Ba}=0$, which is just the conservation law for the EMA currents \cite{Lu15}.

To compare the above result with Lorentz gravity, define the canonical EM and 4d AM tensors
\begin{equation}\label{Vba}
V_b{}^a=V_A{}^{Ba}(D_b\xi^A)(2\xi_B/l^2),
\end{equation}
\begin{equation}\label{Vbca}
V_b{}^{ca}=V_A{}^{Ba}(D_b\xi^A)(D^c\xi_B),
\end{equation}
and, in the same way, the inherent EM tensor and spin tensor
\begin{equation}\label{tauba}
\tau_b{}^a=\tau_A{}^{Ba}(D_b\xi^A)(2\xi_B/l^2),
\end{equation}
\begin{equation}\label{spin}
\tau_b{}^{ca}=\tau_A{}^{Ba}(D_b\xi^A)(D^c\xi_B).
\end{equation}
It can be verified that
\begin{equation}\label{taubaG}
V_b{}^a=\Sigma_b{}^a+\tau_b{}^a,\ \ V_b{}^{ca}=\tau_b{}^{ca},
\end{equation}
and Eq. (\ref{dVG}) is equivalent to
\begin{equation}\label{dV=Vtau}
(\nabla_a+S_a)V_b{}^a=V_c{}^aS^c{}_{ba}+\tau_c{}^{da}R^c{}_{dba},
\end{equation}
\begin{equation}\label{dtau=V}
(\nabla_a+S_a)\tau_{bc}{}^a=-V_{[bc]},
\end{equation}
where $\nabla_a$ is the metric-compatible derivative with torsion, and $S_a=S^c{}_{ac}$.
The above equations have the same form as the EMA conservation equations in Lorentz gravity \cite{Kibble,Hehl76}.
They are also similar to the EM and hypermomentum conservation equations in metric-affine gravity
\cite{Hehl95,MAG}. The difference between dS gravity
and Lorentz gravity (or metric-affine gravity) is rooted in the difference between the dS group and the Lorentz
group (or the general linear group). The matrix dimension of the Lorentz group (or the general linear
group) is equal to the spacetime dimension, while that of the dS group is larger than the spacetime dimension.
As a result, one should introduce the higher-dimensional object $\xi^A$ in the Lagrangian function (\ref{SM}) to
describe the dS invariance; while in Lorentz gravity (or metric-affine gravity), $\xi^A$ is absent.
Also, in dS gravity, the conservation equations (\ref{dV=Vtau})--(\ref{dtau=V})
can be unified in a higher-dimensional conservation equation (\ref{dVG}), which is absent in Lorentz gravity
(or metric-affine gravity). Moreover, in Lorentz gravity (or metric-affine gravity), $V_b{}^a$ in Eqs.
(\ref{dV=Vtau})--(\ref{dtau=V}) is equal to the orbital EM tensor $\Sigma_b{}^a$; while in dS gravity,
it differs from $\Sigma_b{}^a$ by $\tau_b{}^a$, which is defined in Eq. (\ref{tauba}) by the higher-dimensional
object $\tau_A{}^{Ba}$.

To find the observational effects of $\tau_b{}^a$, let us consider a dust fluid with dS spin,
i.e., a fluid with $\Sigma_{ab}=\rho U_aU_b$ and $\tau_A{}^{Ba}\neq0$, where $\rho$ is the rest energy density, and
$U^a$ is the 4-velocity of the fluid particle. Because $\tau_A{}^{Ba}\neq0$, $\tau_b{}^a$ and
$\tau_b{}^{ca}$ are generally nonzero. Assume that the torsion vanishes, then Eq. (\ref{dV=Vtau}) becomes
\begin{equation}\label{dust1}
\nabla_a(\rho U_bU^a+\tau_b{}^a)=\tau_c{}^{da}R^c{}_{dba}.
\end{equation}
Multiplying the above equation by $U^b$ yields
\begin{equation}\label{dust2}
-U^a\nabla_a\rho-\rho\nabla_aU^a+U^b\nabla_a\tau_b{}^a=\tau_c{}^{da}R^c{}_{dba}U^b.
\end{equation}
Substitution of Eq. (\ref{dust2}) into Eq. (\ref{dust1}) leads to
\begin{equation}
\rho U^a\nabla_aU_b=-h^e{}_b(\nabla_a\tau_e{}^a-\tau_c{}^{da}R^c{}_{dea}),
\end{equation}
where $h_{eb}=g_{eb}+U_eU_b$. Obviously, both $\tau_b{}^a$ and $\tau_b{}^{ca}$
contribute to the derivation of $ U^a\nabla_aU_b$ from zero. Hence $\tau_b{}^a$ might be observed
by its contribution to the derivation of the dust particle's world line from a geodesic.
This contribution should be very small, because according to Eq. (\ref{tauba}), $\tau_b{}^a$
is proportional to $l^{-1}\sim\Lambda^{1/2}$ , where $\Lambda=3/l^2$ is the cosmological constant.

Now we turn from the matter field to the gravitational field, with the action integral
$S_G$ and Lagrangian function $\mathscr{L}_G$ given by Eq. (\ref{SG}).
The bundle isomorphisms $\widetilde{\phi}_t$ induce the transformations below:
\begin{eqnarray}\label{gaugedSG}
\Omega\rightarrow\Omega_t=\phi_t[\Omega],\ \
\xi^A\rightarrow \xi_t^A=g_t^A{}_B\phi_{t*}\xi^B,\nonumber\\
D_a\xi^A\rightarrow (D_a\xi^A)_t=g_t^A{}_B\phi_{t*}(D_a\xi^B),\nonumber\\
{\cal F}^A{}_{Bab}\rightarrow{\cal F}_t^A{}_{Bab}=
g_t^A{}_C(\phi_{t*}{\cal F}^C{}_{Dab})(g_t^{-1})^D{}_B,
\end{eqnarray}
which imply
\begin{eqnarray}
\delta\xi^A=B^A{}_B\xi^B-v^aD_a\xi^A,\nonumber\\
\delta D_a\xi^A=B^A{}_BD_a\xi^B-v^bD_bD_a\xi^A-(D_b\xi^A)\mathring{\nabla}_av^b,\nonumber\\
\delta{\cal F}^A{}_{Bab}=[B,{\cal F}_{ab}]^A{}_B-v^cD_c{\cal F}^A{}_{Bab}
-{\cal F}^A{}_{Bcb}\mathring{\nabla}_av^c-{\cal F}^A{}_{Bac}\mathring{\nabla}_bv^c.
\end{eqnarray}
On account of the invariance of $S_G$ under Eq. (\ref{gaugedSG}),
$\delta\mathscr{L}_G=-v^a\mathring{\nabla}_a\mathscr{L}_G$.
Substitution of the above results into the chain rule for $\delta\mathscr{L}_G$ leads to
\begin{equation}\label{dsigma-hat}
\mathring{\nabla}_a\hat{\Sigma}_b{}^a=-\hat{\tau}_A{}^{Ba}{\cal F}^A{}_{Bab}
+\frac{\partial\mathscr{L}_G}{\partial\xi^A}D_b\xi^A+
\frac{\partial\mathscr{L}_G}{\partial D_a\xi^A}D_bD_a\xi^A,
\end{equation}
\begin{equation}\label{dtau-hat}
D_a\hat{\tau}_{AB}{}^a=-\left(\frac{\partial\mathscr{L}_G}{\partial\xi^{[A}}\xi_{B]}
+\frac{\partial\mathscr{L}_G}{\partial D_a\xi^{[A}}D_a\xi_{B]}\right),
\end{equation}
\begin{equation}\label{dxi-F}
\frac{\partial\mathscr{L}_G}{\partial D_a\xi^A}D_b\xi^A
+2\frac{\partial\mathscr{L}_G}{\partial{\cal F}^A{}_{Bac}}{\cal F}^A{}_{Bbc}=0,
\end{equation}
where the Bianchi identity $D_{[c}{\cal F}^A{}_{|B|ab]}=0$, and
the arbitrariness of $B^A{}_B$, $v^a$ and $\mathring{\nabla}_av^b$ at any given point is used,
\begin{eqnarray}\label{sigma-hat}
\hat{\Sigma}_b{}^a=-2\frac{\partial\mathscr{L}_G}{\partial{\cal F}^A{}_{Bac}}{\cal F}^A{}_{Bbc}
+\mathscr{L}_G\delta^a{}_b
\end{eqnarray}
is the orbital EM tensor of gravity, and
\begin{equation}\label{tau-hat}
\hat{\tau}_A{}^{Ba}=2D_b\left(\frac{\partial\mathscr{L}_G}{\partial{\cal F}^A{}_{Bab}}\right)
\end{equation}
is the dS-covariant spin current of gravity. The remaining analyses are similar to
those of the matter field, and one may refer to Eqs. (\ref{sigmaABG})--(\ref{dtau=V}).
It is noteworthy that, unlike the matter field case, there exists the inherent EM tensor for the
gravitational field both in dS gravity and Lorentz gravity.

\section{Conservation Law B}

The total (matter plus gravity) EMA currents cannot be given by the conservation law A in the
last section, because $V_A{}^{Ba}+\kappa\hat V_A{}^{Ba}$ should be equal to zero due to the gravitational
field equation, where $\hat  V_A{}^{Ba}=\delta S_G/\delta\Omega^A{}_{Ba}$.
In order to find the definitions of the total EMA currents, we turn to
the coupling system of the matter field and the gravitational field,
with the action integral $S$ and Lagrangian function $\mathscr{L}$ as follows:
\begin{equation}\label{S}
S=\int_\Omega\mathscr{L}\epsilon,\ \
\mathscr{L}=\mathscr{L}(\psi, D_a\psi,c.c., \xi^A, D_a\xi^A, {\cal F}^A{}_{Bab}),
\end{equation}
where $\mathscr{L}=\mathscr{L}_M+\kappa\mathscr{L}_G$.
The bundle isomorphisms $\widetilde{\phi}_t$ induce the transformations below:
\begin{eqnarray}\label{gaugedS}
\Omega\rightarrow\Omega_t=\phi_t[\Omega],\ \
\psi\rightarrow\psi_t=T(g_t)\phi_{t*}\psi,\ \
\xi^A\rightarrow \xi_t^A=g_t^A{}_B\phi_{t*}\xi^B,
\nonumber \\
\Omega^A{}_{Ba}\rightarrow\Omega_t^A{}_{Ba}
=g_t^A{}_C(\phi_{t*}\Omega^C{}_{Da})(g_t^{-1})^D{}_B
+g_t^A{}_C d_a(g_t^{-1})^C{}_B,
\end{eqnarray}
which imply
\begin{eqnarray}\label{delta-Omega}
\delta\psi=B^A{}_BT_A{}^B\psi-v^aD_a\psi,\nonumber\\
\delta\xi^A=B^A{}_B\xi^B-v^aD_a\xi^A,\nonumber\\
\delta\Omega^A{}_{Ba}=-D_aB^A{}_B+{\cal F}^A{}_{Bab}v^b.
\end{eqnarray}
On account of the invariance of $S$ under Eq. (\ref{gaugedS}),
\begin{equation}
\delta(\mathscr{L}\epsilon)=-\mathring{\nabla}_a(\mathscr{L}v^a)\epsilon.
\end{equation}
Suppose that the gravitational field equation is satisfied, then it follows from
the chain rule for $\delta\mathscr{L}$ that
\begin{eqnarray}
\delta(\mathscr{L}\epsilon)=\mathring{\nabla}_a\left(
\frac{\partial\mathscr{L}}{\partial D_a\psi}\delta\psi+c.c.
+2\frac{\partial\mathscr{L}}{\partial{\cal F}^A{}_{Bab}}\delta\Omega^A{}_{Bb}\right.\nonumber\\
\left.+\frac{\partial\mathscr{L}}{\partial D_a\xi^A}\delta\xi^A
+\mathscr{L}(D^a\xi_A)\delta\xi^A\right)\epsilon.
\end{eqnarray}
Combining the above two results leads to the conservation law B: $\mathring{\nabla}_aJ^a=0$, where
\begin{eqnarray}\label{J}
J^a&=&\frac{\partial\mathscr{L}}{\partial D_a\psi}\delta\psi+c.c.
+2\frac{\partial\mathscr{L}}{\partial{\cal F}^A{}_{Bab}}\delta\Omega^A{}_{Bb} \nonumber\\
&~&+\frac{\partial\mathscr{L}}{\partial D_a\xi^A}\delta\xi^A
+\mathscr{L}(D^a\xi_A)\delta\xi^A+\mathscr{L}v^a.
\end{eqnarray}
Substituting Eq. (\ref{delta-Omega}) into the above equation yields
\begin{eqnarray}\label{J2}
J^a&=&B^A{}_B\frac{\delta S}{\delta\Omega^A{}_{Ba}}
+\mathring{\nabla}_b\left(2B^A{}_B\frac{\partial\mathscr{L}}{\partial{\cal F}^A{}_{Bba}}\right)\nonumber\\
&=&\mathring{\nabla}_b\left(2B^A{}_B\frac{\partial\mathscr{L}}{\partial{\cal F}^A{}_{Bba}}\right),
\end{eqnarray}
where Eqs. (\ref{dpsi-dxi}) and (\ref{dxi-F}) are used, and
\begin{eqnarray}
\frac{\delta S}{\delta\Omega^A{}_{Ba}}&=&\frac{\partial\mathscr{L}}{\partial D_a\psi}T_A{}^B\psi+c.c.
+2D_b\frac{\partial\mathscr{L}}{\partial{\cal F}^A{}_{Bab}}\nonumber\\
&~&+\left(\frac{\partial\mathscr{L}}{\partial D_a\xi^{[A}}\xi_{C]}
+\mathscr{L}(D^a\xi_{[A})\xi_{C]}\right)\eta^{BC},
\end{eqnarray}
which vanishes due to the gravitational field equation. The current (\ref{J2})
is the 5d dS-invariant conserved AM current with respect to $\{\widetilde{\phi}_t\}$.
Note that $v^a$ does not appear in the final expression of $J^a$, and thus the
one-parameter group of horizontal gauge transformations (which may be interpreted as
spacetime diffeomorphisms) does not correspond to any conserved current.
Indeed, for a one-parameter group of horizontal gauge transformations,
Eq. (\ref{horizontal}) holds, and so $B^A{}_B$ defined in Eq. (\ref{B}) is equal to
zero, and hence the conserved current $J^a=0$ on account of Eq. (\ref{J2}).

In dS SR, the dS-invariant AM current with respect to the dS symmetry characterized by
$B^A{}_B$ is equal to \cite{Lu15}
\begin{equation}
J_{\rm SR}^{\,a}=B^A{}_BV_A{}^{Ba}.
\end{equation}
It is seen that $B^\alpha{}_4$ characterizes the symmetry corresponding to the EM current,
and $B^\alpha{}_\beta$ characterizes the symmetry corresponding to the 4d AM current.
As a result, the total (matter plus gravity) EM current can be defined as
\begin{equation}\label{JEM}
J^a(B^\alpha{}_4)=\mathring{\nabla}_b\left(
4B^\alpha{}_4\frac{\partial\mathscr{L}}{\partial{\cal F}^\alpha{}_{4ba}}\right),
\end{equation}
and the total 4d AM current can be defined as
\begin{equation}\label{JAM}
J^a(B^\alpha{}_\beta)=\mathring{\nabla}_b\left(
2B^\alpha{}_\beta\frac{\partial\mathscr{L}}{\partial{\cal F}^\alpha{}_{\beta ba}}\right).
\end{equation}
They are gauge dependent, but constitute the gauge-independent 5d AM current (\ref{J2}).
In dS gravity, there exist the Lorentz gauges, such that \cite{Stelle,Lu13,Lu14}
\begin{equation}
\xi^A=\left(
\begin{array}{cc}
0_{4\times1}\\l
\end{array}
\right),\ \
D_a\xi^A=\left(
\begin{array}{cc}
e^\alpha{}_a\\0
\end{array}
\right),
\end{equation}
\begin{equation}
\Omega^{A}{}_{Ba}=\left(
\begin{array}{cc}
\Gamma^{\alpha}{}_{\beta a}&l^{-1}e^{\alpha}{}_{a}\\
-l^{-1}e_{\beta a}&0
\end{array}
\right),
\end{equation}
\begin{equation}
\mathcal {F}^{A}{}_{Bab}=\left(
\begin{array}{cc}
R^{\alpha}{}_{\beta ab}-l^{-2}e^{\alpha}{}_{a}\wedge e_{\beta b}
&l^{-1}S^{\alpha}{}_{ab}\\
-l^{-1}S_{\beta ab}&0
\end{array}
\right),
\end{equation}
where $0_{4\times1}=(0,0,0,0)^{\rm T}$, $\{e_\alpha{}^a\}$ is a tetrad field,
$\alpha=0,1,2,3$ are tetrad indices, $e^\alpha{}_ae_\beta{}^a=\delta^\alpha{}_\beta$,
$e_{\beta a}=\eta_{\alpha\beta}e^\alpha{}_a$, $\eta_{\alpha\beta}$ is the Minkowski metric,
$\Gamma^{\alpha}{}_{\beta a}$ is the metric-compatible connection 1-form of spacetime,
$S^{\alpha}{}_{ab}=d_a e^{\alpha}{}_b+\Gamma^{\alpha}{}_{\beta
a}\wedge e^{\beta}{}_{b}$ is the torsion 2-form, and $R^{\alpha}{}_{\beta ab}
=d_a\Gamma^{\alpha}{}_{\beta b}+\Gamma^{\alpha}{}_{\gamma
a}\wedge \Gamma^{\gamma}{}_{\beta b}$ is the curvature 2-form.
The dS-invariant current (\ref{J2}) becomes a Lorentz-invariant current
\begin{eqnarray}\label{JLG}
J^a=\mathring{\nabla}_b\left(2B^\alpha{}_\beta\frac{\partial\mathscr{L}}{\partial R^\alpha{}_{\beta ba}}
+2V^\alpha\frac{\partial\mathscr{L}}{\partial S^\alpha{}_{ba}}\right)
\end{eqnarray}
in the Lorentz gauges, which recovers the conserved current for Lorentz gravity \cite{Obukhov06,Obukhov08},
where $V^\alpha=B^\alpha{}_4l$. In Eq. (\ref{JLG}), the $B^\alpha{}_\beta$ term is the AM current
corresponding to Lorentz rotations, while the $V^\alpha$ term is the EM current corresponding to the
translations defined by the dynamical connection in Lorentz gravity \cite{Hehl76,Obukhov06}.
A drawback of the above definitions is that the Komar EM current
$J_K^a=\mathring{\nabla}_b(2\kappa\mathring{\nabla}^{[b}V^{a]})$ \cite{Komar} cannot be
explained as an EM current in the Einstein--Cartan gravity, where
$V^a=V^\alpha e_\alpha{}^a$. Instead, $J_K^a$ should be interpreted as an AM current with
\begin{equation}
B^\alpha{}_\beta=
(\mathring{\nabla}_{[a}V_{b]})e^{\alpha a}e_\beta{}^b.
\end{equation}
The conserved current (\ref{JLG}) is also similar to the $GL(4,\mathbb{R})$-invariant
conserved current in metric-affine gravity \cite{MAG}, where $GL(4,\mathbb{R})$ is
the general linear group. As mentioned before, the difference between dS gravity and Lorentz
gravity (or metric-affine gravity) is rooted in the higher-dimensional feature of the dS group.
To describe the dS invariance, one should introduce the higher-dimensional object $\xi^A$ in the
Lagrangian function (\ref{S}); while in Lorentz gravity (or metric-affine gravity), $\xi^A$ is absent.
Also, in dS gravity, the EM and AM currents (\ref{JEM})--(\ref{JAM})
can be unified in a higher-dimensional current (\ref{J2}), which is absent in Lorentz gravity
(or metric-affine gravity).

\section{Remarks}

In this paper, two kinds of conservation laws are discussed in dS gravity.
The first kind is a differential equation (\ref{dVG}) for the dynamical
current $V_A{}^{Ba}=\delta S_M/\delta\Omega^A{}_{Ba}$.
It is resulted from the diffeomorphism and dS symmetries of the theory.
The dS-covariant current $V_A{}^{Ba}$ unites the EM and spin tensors, while
Eq. (\ref{dVG}) unites two differential equations (\ref{dV=Vtau})--(\ref{dtau=V})
which have the same form as the conservation equations in Lorentz gravity.
Moreover, the EM tensor is found to be containing an inherent part, which
might be detected by its contribution to the derivation of the dust particle's
world line from a geodesic.
The second kind of conservation law presents a dS-invariant conserved current (\ref{J2})
with respect to each one-parameter group of dS rotations. The dS-invariant current
unites the total EMA currents (\ref{JEM})--(\ref{JAM}) for the coupling system
of matter and gravity. Also, it is shown that the conserved current with respect to
the diffeomorphism symmetry is equal to zero, in other words, the diffeomorphism symmetry does
not lead to any conserved current, and so the Noether's theorem does not completely
apply to dS gravity.

Base on these discussions, we give a summary on some differences between dS gravity
and Lorentz gravity as follows. Firstly, in dS gravity, one should introduce
the higher-dimensional object $\xi^A$ in the Lagrangian function to illustrate how
the system is dS invariant; while in Lorentz gravity, $\xi^A$ is absent.
Secondly, in dS gravity, the orbital AM current can be well defined by using $\xi^A$,
see Eq. (\ref{sigmaABG}); while in Lorentz gravity, it should be equal to zero.
Thirdly, in dS gravity, the EM and AM conservation equations (\ref{dV=Vtau})--(\ref{dtau=V})
can be unified in a higher-dimensional conservation equation (\ref{dVG});
while in Lorentz gravity, they cannot be unified. Fourthly, in dS gravity, the EM
and AM currents (\ref{JEM})--(\ref{JAM}) can be unified in a higher-dimensional current (\ref{J2});
while in Lorentz gravity, they cannot be unified. Fifthly, in dS gravity, the inherent EM tensors
of both the matter field and the gravitational field are nonzero; while in Lorentz gravity,
the inherent EM tensor of the matter field vanishes, but that of the gravitational field is nonzero.
These differences are rooted in the higher-dimensional feature of the dS group.

\section*{Acknowledgments}
I would like to thank the late Prof. H.-Y. Guo, and Profs. C.-G. Huang, Z.-B. Li,
X.-W. Liu and M. Li for their help
and the useful discussions. The project is funded by the China Postdoctoral Science Foundation
(Grant No. 2015M572393), and the Fundamental Research Funds for the Central
Universities (Grant No. 161gpy49) at Sun Yat-sen University.

\end{document}